\documentclass[AEJ]{AEA}

\usepackage{natbib}
\usepackage{comment}
\usepackage{graphicx}
\usepackage{rotating}
\usepackage{lscape}
\usepackage{geometry}
\usepackage{hyperref}
\hypersetup{
colorlinks = true,
linkcolor = cyan,
citecolor = magenta,
urlcolor = blue,
}

\draftSpacing{1.5}

\begin{document}

\title{Rise of the Kniesians: The professor-student network of Nobel laureates in economics}
\shortTitle{Rise of the Kniesians}
\author{Richard S.J. Tol\thanks{Tol: Department of Economics, University of Sussex, BN1 9SL Falmer, United Kingdom, r.tol@sussex.ac.uk; Institute for Environmental Studies, Vrije Universiteit Amsterdam, The Netherlands; Department of Spatial Economics, Vrije Universiteit Amsterdam, The Netherlands; Tinbergen Institute, Amsterdam, The Netherlands; CESifo, Munich, Germany; Payne Institute of Earth Resources, Colorado School of Mines, Golden, CO, USA. Richard Blundell, Ernst Fehr, Benjamin Friedman, Jerry Hausman, David Hendry, Edward Lazear, Richard Levin, John Moore, Gerhard Orosel and Larry Summers kindly answered questions about their ancestry. Two anonymous referees, Irwin Collier, Peter Dolton and Stephen Zeff had helpful information and comments.}}
\date{\today}
\pubMonth{Month}
\pubYear{Year}
\pubVolume{Vol}
\pubIssue{Issue}
\JEL{A14, B20, B31}
\Keywords{social network; professor-student relationship; Nobel Prize}

\begin{abstract}
The paper presents the professor-student network of Nobel laureates in economics. 82 of the 87 Nobelists belong to one family tree. The remaining 5 belong to 3 separate trees. There are 376 men in the graph, and 5 women. Karl Knies is the central-most professor, followed by Wassily Leontief. No classical and few neo-classical economists have left notable descendants. Harvard is the central-most university, followed by Chicago and Berlin. Most candidates for the Nobel prize belong to the main family tree, but new trees may arise for the students of Terence Gorman and Denis Sargan.
\end{abstract}

\maketitle

\begin{quote}
    "It is Knies, however, whom I am glad to acknowledge, more than any other one man, as My Master." \citep{Ely1938}
\end{quote}

\section{Introduction}
The highest accolade in economics, the Sveriges Riksbank Prize in Economic Sciences in Memory of Alfred Nobel and its winners attract extraordinary attention, including how different Laureates relate to each other. Besides the gossip, particular attention has been paid to schools of thought and influential institutions \citep{McCarty2000, Vane2005, Breit2009, Boettke2012, Chan2012, Claes2013, Spencer2015, Solow2015, Offer2016}. Although anecdotes of who studied with whom are widely known, no one has comprehensively mapped the professor-student relationships of all Nobel Laureates in economics. This paper does just that.

I show that most Nobelists are related, often closely.\footnote{\citet{Zuckerman1996} and \citet{Chan2015} show that Nobelists in other disciplines cluster too.} The 2017 Laureate, Richard H. Thaler, is typically described as an outsider in the press \citep{Appelbaum2017, Avent2017, Harford2017, Partington2017, Vitasek2017}\textemdash but that may be a mischaracterisation of someone with a PhD from the University of Rochester, Gary S. Becker and Robert E. Lucas as academic uncles, and James J. Heckman as an academic cousin. William J. Baumol and his student Lionel W. McKenzie were not honored by a trip to Stockholm, inexplicably so according to some \citep{Minniti2016, Weintraub2011}. Actually, Baumol is the academic brother of John R. Hicks, and uncle to Franco Modigliani, Amartya Sen and Joseph E. Stiglitz. Paul M. Romer is McKenzie's grandstudent. Some argue that Margaret G. Reid should have shared the Nobel Prize with Gary S. Becker \citep{Yi1996}. Whatever the merits of this argument, a lack of connections was not part of it: Reid is Ronald Coase' first cousin once removed, she had a PhD from the University of Chicago, and taught there for many years. A comprehensive map of all Laureates, including a way to connect those that could or should have won, is useful to dispel myths and ugly rumours.

You can interpret the close relationships between Nobelists as a manifestation of the clustering of quality: The best professors congregate in the best schools \citep{Ellison2013}, reinforce each other \citep{Azoulay2010, Borjas2012, BOSQUET2017, Oyer2006}, and select the best students \citep{Athey2007}. You can also see this as a manifestation of the feudal nature of academia and the nepotism implied \citep{COMBES2008, Hamermesh2003, Laband1994, Medoff2003}. \citet{Zuckerman1996} points out that Nobelists have the right to nominate, and put their proteges forward. She argues, though, that matching on quality, training, and competition between the students of Nobel Laureates are more important in explaining the success of Nobel students.

Tracing the Nobel network over time illustrates key features of the development of economics as a discipline. The Scandinavian branches of the network were ascendant for a while, but key people stayed put in or returned to Norway and Sweden and lost global influence, unlike the Dutch who moved to the USA and kept winning. The French branch moved to the USA too, and withered. The current network is dominated by the UK and the USA, with competition between East Coast and Great Lakes, and deep roots in Austria, Germany and France uniting many. Intriguingly, none of the great classical economists\textemdash Smith, Say, Ricardo, Malthus, Mill, Marx\textemdash and pre-classical economists\textemdash Ibn Khaldun, Petty, Quesnay, Cantillon, Turgot, Galiani\textemdash is connected to the Nobelists, and only one of the neo-classical revolutionaries: Carl Menger\textemdash Marshall, Walras, Jevons, Pareto and Pigou made enduring contributions to economic thought, but only through their writings, not through their students.\footnote{John Maynard Keynes was taught and mentored by Marshall and Pigou, but was supervised by W.E. Johnson and A.N. Whitehead.} In those days, apparently, economics was not yet a discipline that could be taught to young scholars, or at least the leaders of the profession did not see it this way.

This is specific to economics: Scholars in the humanities and natural sciences can trace their ancestry to great names of times past, and indeed many economists descend from famous non-economists\textemdash including people whose work we use, such as the Bernouillis, Gauss, Lagrange, Lyapunov and Pearson, and people whose work is less obviously relevant, such as Bohr, Erasmus, Heisenberg, Luther and Maxwell. In other words, economics was taken over by people from other disciplines, much like economists now work on subjects that traditionally were the exclusive domain of political scientists, anthropologists, psychologists and biologists.

Representing the network as a graph and adding degree-granting institutions, I identify central professors and schools. The results may be surprising. A relatively obscure member of the German Historical School, Karl Knies emerges as the central-most thesis advisor. Knies is not nearly as famous as his students Eugen B{\"o}hm von Bawerk \citep{VonMises2016}, John Bates Clark \citep{Homan1927, Leonard2003}, and Richard T. Ely \citep{Rader1966, Thies2010}. Berlin, G\"{o}ttingen and Vienna are the top 10 of central-most schools, with more obvious candidates such as Harvard and Chicago.

I also consider the likely candidates for future Nobel Prizes, revealing that the current network is likely to produce future winners. New networks may emerge as well, particularly around the students, grandstudents and great-grandstudents of Terence Gorman.

The paper proceeds as follows. Data and methods are presented in the next section. Section 3 discusses the results. Section 4 concludes.

\section{Data and methods}

\subsection{Data}
The list of Nobel laureates was taken from \href{https://ideas.repec.org/nobel.html}{IDEAS/RePEc}. The list of Nobel candidates is Clarivate's 2020 list of \href{https://clarivate.com/hall-of-citation-laureates/}{Citation Laureates}. Speculation is rife in the run-up to the Nobel Prize announcement and lists of candidates abound. The Clarivate list is based on objective criteria\textemdash citation numbers in economics journals\textemdash and has a reasonable prediction record. It largely overlaps with other lists.

Data on ancestry and final degree were gathered from and stored on \href{https://academictree.org/}{AcademicTree.org}, a collaborative tool for building an academic genealogy for all disciplines \citep{David2012}. The economics tree\footnote{I follow the data source and use the colloquial expression ''tree'' in this paper. Strictly, this is a directed acyclic graph or polytree.} contains over 21,000 economists, the overall tree almost 700,000 academics. Where AcademicTree was incomplete, I added data from the \href{https://www.genealogy.math.ndsu.nodak.edu/}{Mathematics Genealogy Project}, which actually includes many economists, and from \href{https://genealogy.repec.org/}{RePEc Genealogy}, which focuses on economists but is limited in its historical depth. If all three sources fell short, I used Wikipedia, studied CVs, obituaries and acknowledgements in early papers, and dug up the occasional thesis. When all that failed, I contacted people or their students directly. The results can be found on \href{https://academictree.org/}{AcademicTree.org}. The data can be amended where needed and extended to include the reader's favorite economists.

The data are not perfect. Formal professor-student relationships are recorded, but may be less important than informal mentoring.\footnote{\href{https://academictree.org/econ/tree.php?pid=753646}{Friedrich von Wieser} is one example. He co-advised Friedrich von Hayek with Carl Menger and Ludwig von Mises, but as Menger was both Mises' and Wieser's advisor, Menger takes most of the numerical credit for Hayek's success. Wieser influenced Mises and Joseph Schumpeter, but his brother-in-law Eugen von B\"{o}hm-Bawerk was their official advisor. The influence metric presented below does not do justice to Wieser's contribution.} Some professors are more important than others.\footnote{I had three PhD advisers. One taught me statistics, one taught me office politics, and one taught me nothing before I finished my PhD but much after. I learned economics from people who were not on my committee.} The formalization of research training is a recent development, and different countries made this transition at different times. In the generation of professors now retiring in the UK and the Netherlands, it was not uncommon to not have a PhD. Germany and Austria have the \textit{Habilitation}, a degree after the PhD. France used to have that too. An Italian doctorate is often seen as the degree to enter a PhD in England or the USA. These norms have changed over the centuries that are spanned by the data. The edges in the graph are not uniform. Some data are missing, and some people were self-taught\textemdash no distinction is made. For earlier generations, historians have focused on prominent scholars, their professors and their students. Some universities have excellent online records, other universities closely guard graduation data.

The professors in the tree presented here are, in most cases, the formal PhD advisors to the students in the tree; often two per student, but one advisor or three are not uncommon. Where appropriate, the advisor of the Habilitation or Master's thesis is either added to or used instead of the PhD advisor. In a few cases, an informal mentor is used.

The data are taken at face value. No probabilities are assigned, or fuzzy memberships. A limited sensitivity analysis is shown below. Data and algorithms are in the public domain for anyone to add, alter or perform their preferred sensitivity analysis.

\subsection{Methods}
Data were transferred to Matlab for visualization and analysis. Five generations were included, unless graphs could be connected by including more distant ancestors, in which case the closest common ancestor was included. Without these two restrictions, the network would be dominated by the intricate relationships of non-economists of times long past and the somewhat speculative lineage of Jesus Christ. Data are interpreted as a directed acyclic graph\textemdash because in most cases students learn more from their professors than the other way around\textemdash with unweighted edges\textemdash because we do not know, in most cases, which advisor was more influential. I use a minor modification of an outcloseness measure $c^A(i)$ for network centrality:
\begin{equation}
\label{eq:arithmean}
    c^A(i) = \left ( \frac{A_i}{N-1} \right )^2 \frac{1}{\sum_j C_{i,j}}
\end{equation}
where $A_i$ is the number of \textit{Nobel} nodes that can be reached from node $i$, $N$ is the number of nodes, and $C_{i,j}$ is the distance from node $i$ to any \textit{Nobel} node. If the word \textit{Nobel} is dropped, this reverts to the measure of outcloseness as defined by \citet{Bavelas1950} for a connected graph and extended, in a seemingly \textit{ad hoc} but not unreasonable way, by Matlab to unconnected graphs by multiplication with the square of the fraction of nodes reachable.

In a connected graph, Equation (\ref{eq:arithmean}) is the \textit{arithmetic} mean of the distance to all other nodes. \citet{Marchiori2000} propose the \textit{harmonic} mean as a measure of distance:\footnote{Earlier, \citet{GilSchmidt1996} suggested its inverse as a measure of closeness.}
\begin{equation}
\label{eq:harmmean}
    c^H(i) = (N-1)\sum_j \frac{1}{C_{i,j}}
\end{equation}
The key advantage of the harmonic mean is that it applies to connected as well as unconnected graphs\textemdash for unreachable nodes $C_{i,j} = \infty$. In Equation (\ref{eq:harmmean}), the number of unreachable nodes is penalized linearly; in Equation (\ref{eq:arithmean}), the penalty is quadratic. In this context, when distance is restricted to distance to a Nobel laureate, the harmonic mean has the additional advantage that proximity is emphasized at the expense of distal relationships. I therefore use the harmonic mean distance as my key measure of centrality, and the arithmetic mean as a robustness check.

Note that outcloseness on a polytree measures ancestry. Two students of the same professor are not close. The professor is close to both of them, but they are not close to each other.\footnote{An undirected graph would show that academic siblings are close. However, as professors frequently co-advise new students with their former students, an undirected graph would have cycles.}

For Nobel candidates, what matters is their distance to the graph, rather than the graph's distance to them. This is an incloseness measure, rather than an outcloseness one. It is readily computed by replacing $C_{i,j}$ by $C_{j,i}$ in Equation (\ref{eq:harmmean}).

I measure the change in the network over time by the Graph Edit Distance \citep{SanfeliuFu1983}. The edit distance between graph $A$ and $B$ equals the number of nodes and edges that need to be added to (or removed from) $A$ to make $A$ isomorphic to $B$. Typically, computing the Graph Edit Distance is a n-p hard problem, but in this case the graph for the previous year is a subgraph of the graph for the current year. Indeed, the algorithm that builds the graph for year $t$ uses the graph for year $t-1$ as its starting point. This also means that the edits make the graphs identical, rather than just isomorphic.

Besides the changes in the overall graph over time, I am also interested in changes in the centrality of individuals. While the centrality measures defined above have a cardinal interpretation at any point in time, this is not true for changes in centrality.\footnote{This is easily seen. Consider a graph. Add a node with a single edge to the central-most node. Add $N$ unconnected nodes. If $N$ is sufficiently large, centrality falls numerically, even though the central-most node has become more central.} I therefore use changes in the centrality \textit{rank} to assess changes over time.

\section{Nobelists}

\href{https://academictree.org/econ/tree.php?pid=740232}{Ragnar Frisch}, \href{https://academictree.org/econ/tree.php?pid=23345}{Jan Tinbergen}\footnote{Jan's biological brother, Niko, won the Nobel Prize for Physiology in 1973.} and \href{https://academictree.org/econ/tree.php?pid=162776}{Paul A. Samuelson} all start their own graphs. That is, they do not share an ancestor. \href{https://academictree.org/geography/tree.php?pid=299692}{Simon S. Kuznets} and Samuelson do share an ancestor in Christian Gottlob Heyne, an 18\textsuperscript{th} century classicist and archaeologist at the University of G{\"o}ttingen, Germany. \href{https://academictree.org/econ/tree.php?pid=12113}{Kenneth J. Arrow} starts his own graph.\footnote{Arrow's biological sister was married to Samuelson's biological brother.} \href{https://academictree.org/econ/tree.php?pid=742628}{John R. Hicks} was a grandstudent of Carl Menger, who was also a great-grandprofessor of Samuelson. \href{https://academictree.org/econ/tree.php?pid=12124}{Wassily Leontief} was advisor of Samuelson, who is thus the first second-generation Nobelist. Leontief was the first to win the Nobel Prize after his student. \href{https://academictree.org/econ/tree.php?pid=225561}{F.A. Hayek}, Hicks' adviser, was the second. \href{https://academictree.org/econ/tree.php?pid=742629}{Gunnar Myrdal}\footnote{Myrdal's wife, Alva Reimer, won the Nobel Peace Prize in 1982.} is a distant descendant of Pierre Varignon, an 18\textsuperscript{th} century mathematician at the Royal Academy in Paris, France; Arrow also descends from Varignon. After five years, there are nine Nobel laureates, five of whom are part of a single family tree; there is one tree with two Nobelists (Arrow, Myrdal); and two with one (Frisch, Tinbergen).

\href{https://neurotree.org/neurotree/tree.php?pid=742661}{Leonid Kantorovich} and Leontief are distant descendents of Johann Friedrich Pfaff, a 19\textsuperscript{th} century mathematician at the University of Halle-Wittenberg, Germany. \href{https://academictree.org/econ/tree.php?pid=23347}{Tjalling C. Koopmans} was a student of Tinbergen and Hendrik Kramer, who descends from Heyne. Koopmans thus connects the Tinbergen graph to the main one. \href{https://academictree.org/econ/tree.php?pid=738971}{Milton Friedman} was a student of Kuznets, and shares an ancestor with Samuelson in Karl Knies, a 19\textsuperscript{th} century economist at the University of Heidelberg, Germany.

\href{https://academictree.org/econ/tree.php?pid=742633}{James E. Meade} starts his own graph. Like Myrdal (and Eli Heckscher), \href{https://academictree.org/econ/peopleinfo.php?pid=742631}{Bertil Ohlin} was a student of Karl Gustav Cassel. This completes the Swedish branch of the graph\textemdash in contrast with the Dutch branch, which continues to grow, as key people moved to the USA.

\href{https://academictree.org/econ/tree.php?pid=11024}{Herbert A. Simon} was a student of Henry Schultz, like Friedman was. \href{https://academictree.org/econ/tree.php?pid=740768}{Theodore W. Schultz} was a great-grandstudent of Knies. \href{https://academictree.org/history/tree.php?pid=740770}{W. Arthur Lewis} and Hicks are both grandstudents of Edwin Cannan. \href{https://academictree.org/history/tree.php?pid=162773}{Lawrence R. Klein} was Samuelson's student. \href{https://academictree.org/econ/tree.php?pid=739286}{James Tobin} was Joseph Schumpeter's student, like Samuelson, and a grandstudent of Werner Sombart, who with Ladislaus Bortkiewicz advised Leontief.\footnote{A referee points out that Sombart called on Bortkiewicz to check Leontief's mathematics.} \href{https://neurotree.org/neurotree/tree.php?pid=738967}{George J. Stigler} was a great-grandstudent of both Clark and Ely. In 1982, 14 years after the Nobel Memorial Prize in Economic Sciences was first awarded, there is a well-established family tree.

\href{https://academictree.org/econ/tree.php?pid=740273}{Gerard Debreu} starts his own graph. Like Meade, \href{https://academictree.org/econ/tree.php?pid=737466}{Richard Stone} was a student of John Maynard Keynes. He was also a great-grandstudent of Ely and so related to Schultz and Stigler. Stone thus connects the UK and the US graphs. Stone also brings the first female to the graph: Beatrice Potter Webb,\footnote{Beatrice Potter and Beatrix Potter are different people.} who co-founded the London School of Economics.

\href{https://academictree.org/econ/tree.php?pid=739007}{Franco Modigliani} is a grandstudent of Lionel Robbins, one of Hicks' advisers, and a great-grandstudent of Bawerk. Like Stigler, \href{https://academictree.org/econ/tree.php?pid=740272}{James M. Buchanan} was a student of Frank H. Knight. Like Samuelson, \href{https://academictree.org/econ/tree.php?pid=12123}{Robert M. Solow} was a student of Leontief.

Debreu was a post-doc with \href{https://neurotree.org/neurotree/tree.php?pid=740271}{Maurice F.C. Allais}. This completes the French graph. \href{https://academictree.org/econ/tree.php?pid=740231}{Trygve M. Haavelmo} was a student of Frisch. This completes the Norwegian graph.

\href{https://academictree.org/econ/tree.php?pid=740229}{William F. Sharpe} was a great-grandstudent of Harold Hotelling, who was Arrow's adviser. \href{https://academictree.org/econ/tree.php?pid=739189}{Merton M. Miller} was a grandstudent of Ludwig von Mises, Hayek's advisor. \href{https://academictree.org/econ/tree.php?pid=739011}{Harry M. Markowitz} was Jacob Marschak's student, like Modigliani. Markowitz was also a student of Friedman, who was Kuznets' student. Markowitz is thus the first third-generation Nobelist.

Like Stigler and Buchanan, \href{https://academictree.org/econ/tree.php?pid=740017}{Ronald H. Coase} was a student of Knight. \href{https://academictree.org/econ/tree.php?pid=415327}{Gary S. Becker} was a grandstudent of Schultz, the adviser of Friedman and Simon. \href{https://academictree.org/geography/tree.php?pid=299691}{Robert W. Fogel} was Kuznets' student. \href{https://academictree.org/econ/tree.php?pid=299695}{Douglas C. North} was advised by Melvin M. Knight,\footnote{Melvin was a biological brother of Frank.} who was a grandstudent of Henry Moore, like Friedman and Simon.

\href{https://academictree.org/econ/tree.php?pid=739934}{John C. Harsanyi} was a student of Arrow. \href{https://academictree.org/econ/tree.php?pid=739935}{John F. Nash} and \href{https://academictree.org/econ/tree.php?pid=739938}{Reinhard J.R. Selten} are distant descendants of Carl Gauss, and so is Leontief. Selten also descends from Simeon-Denis Poisson as does Hotelling. Selten thus connects the Arrow-Sharpe-Harsanyi/Myrdal-Ohlin graph with the main one.

\href{https://academictree.org/econ/tree.php?pid=738977}{Robert E. Lucas} was a student of Gregg Lewis, like Becker. Lucas is also a great-great-grandstudent of Ely, and so related to many other Nobelists. \href{https://academictree.org/econ/tree.php?pid=738849}{James E. Mirrlees} was a student of Stone. \href{https://academictree.org/econ/tree.php?pid=739862}{William S. Vickrey} was a grandstudent of Edwin Seligman, like Frank Knight. \href{https://academictree.org/econ/tree.php?pid=645972}{Robert C. Merton} was a student of Samuelson. \href{https://academictree.org/econ/tree.php?pid=596630}{Myron S. Scholes} was Miller's student. Eugene F. Fama was Scholes' other adviser. Sixteen years after receiving his prize, Scholes would become the first Nobelist with two Nobel advisers.

\href{https://academictree.org/philosophy/tree.php?pid=175042}{Amartya Sen} was a grandstudent of Robbins, like Modigliani. Joan Robinson was an adviser. She is the second woman in the graph. \href{https://academictree.org/econ/tree.php?pid=22197}{Robert A. Mundell} was a great-grandstudent of James Laughlin, Kuznets' grandprofessor, and of Allyn Abbot Young, Knight's professor. \href{https://academictree.org/econ/tree.php?pid=12121}{Daniel L. McFadden} was a grandstudent of Koopmans and Jakob Marschak, who advised Modigliani. \href{https://academictree.org/econ/tree.php?pid=12122}{James J. Heckman} was a grandstudent of Lewis and Tobin, and a great-grandstudent of Klein. \href{https://academictree.org/econ/tree.php?pid=415994}{George A. Akerlof} was a student of Solow, \href{https://academictree.org/econ/tree.php?pid=738896}{A. Michael Spence} a grandstudent of Leontief. \href{https://academictree.org/econ/tree.php?pid=416057}{Joseph E. Stiglitz} was a student of Robinson, a grandstudent of Robbins and a great-grandstudent of Young. Year after year, the Nobel Prize is awarded to close relatives of previous winners.

\href{https://academictree.org/econ/tree.php?pid=688485}{Vernon L. Smith} was a student of Leontief. \href{https://academictree.org/econ/tree.php?pid=4483}{Daniel Kahneman}'s great-great-great-great-grandprofessor was John Dewey, who was grandprofessor of Kuznets. One of Kahneman's advisers was Susan Ervin-Tripp, the third woman in the graph. \href{https://academictree.org/econ/tree.php?pid=739427}{Robert F. Engle} was a great-great-grandstudent of Wesley Clair Mitchell, Kuznets' professor. \href{https://academictree.org/econ/tree.php?pid=566536}{Clive W.J. Granger} was a great-great-great-grandstudent of Eliakim Moore, Hotelling's grandprofessor. \href{https://academictree.org/econ/tree.php?pid=191388}{Finn E. Kydland} was a student of \href{https://academictree.org/econ/tree.php?pid=415443}{Edward C. Prescott}, a grandstudent of Leontief. They are the first student-professor pair to jointly win. \href{https://academictree.org/econ/tree.php?pid=739288}{Robert J. Aumann} was great-grandstudent of Solomon Lefschetz, who was Nash' grandprofessor. \href{https://academictree.org/econ/tree.php?pid=185525}{Thomas C. Schelling} was Leontief's student, and a grandstudent of Schumpeter, who advised Samuelson and Tobin. \href{https://academictree.org/econ/tree.php?pid=183309}{Edmund S. Phelps} was Tobin's student. \href{https://academictree.org/econ/tree.php?pid=12119}{Eric S. Maskin} and \href{https://academictree.org/econ/tree.php?pid=12118}{Roger B. Myerson} are Arrow's students. \href{https://academictree.org/econ/tree.php?pid=59887}{Leonid Hurwicz} was Koopmans' student and a post-doc with Samuelson. Hurwicz is thus a double third-generation Nobelist. McFadden is a fourth-generation Nobelist, and a double one at that.

\href{https://academictree.org/econ/tree.php?pid=22195}{Paul Krugman} was a grandstudent of Mundell. \href{https://academictree.org/econ/tree.php?pid=519726}{Oliver Williamson} was Simon's student and Stigler's grandstudent. \href{https://academictree.org/polysci/tree.php?pid=195666}{Elinor Ostrom} is the first female Nobelist, and only the fourth woman in the graph. Her grandprofessor was Robert K. Merton\footnote{Robert K. was the biological father of Robert C.}, a descendant of Friedrich Trendelenburg, Dewey's grandprofessor. Like Kahneman before her, Ostrom is part of the main family of Nobelists.

\href{https://academictree.org/econ/tree.php?pid=739107}{Peter A. Diamond} was Solow's student, and \href{https://academictree.org/econ/tree.php?pid=409729}{Dale T. Mortensen} Leontief's grandstudent. \href{https://academictree.org/econ/tree.php?pid=739245}{Christopher A. Pissarides} has his own graph.

\href{https://academictree.org/econ/tree.php?pid=181698}{Thomas J. Sargent} was a great-grandstudent of Arthur Smithies, Schumpeter's student and Schelling's adviser. \href{https://academictree.org/econ/tree.php?pid=188222}{Christopher A. Sims} was Harold S. Houthakker's student, who was Pieter de Wolff's student, who was Tinbergen's student.\footnote{De Wolff is my great-grandprofessor.} \href{https://academictree.org/math/tree.php?pid=177418}{Lloyd S. Shapley} was Alfred Tucker's student, like Nash. \href{https://academictree.org/econ/tree.php?pid=181680}{Alvin E. Roth} descends from Gauss and Poisson. \href{https://academictree.org/econ/tree.php?pid=739191}{Lars Peter Hansen} is Sims' student, \href{https://academictree.org/econ/tree.php?pid=172518}{Robert J. Shiller} Modigliani's. \href{https://academictree.org/econ/tree.php?pid=557291}{Eugene F. Fama} is Miller's student and Scholes' professor. \href{https://academictree.org/econ/tree.php?pid=739109}{Jean Tirole} is Maskin's student, \href{https://academictree.org/econ/tree.php?pid=116843}{Angus Deaton} Stone's. \href{https://academictree.org/econ/tree.php?pid=739108}{Bengt Holmstr{\"o}m} is Robert Wilson's student, like Roth. \href{https://academictree.org/econ/tree.php?pid=175237}{Oliver Hart} is a grandstudent of Solow and Diamond. Like Heckman, \href{https://academictree.org/econ/tree.php?pid=566532}{Richard H. Thaler} is a grandstudent of Lewis, Becker's and Lucas' adviser.

\href{https://academictree.org/econ/tree.php?pid=172510}{William D. Nordhaus} is the third of Solow's students to win the Nobel prize. \href{https://academictree.org/econ/tree.php?pid=519752}{Paul Romer} studied with a Nobel laureate too (Lucas). His other advisor was Jos\'{e} Scheinkman, who is connected with the rest of the tree via Lionel McKenzie, William Baumol and Oskar Morgenstern.

\href{https://academictree.org/econ/tree.php?pid=616505}{Esther Duflo} shared the Nobel Prize with two of her advisers, \href{https://academictree.org/econ/tree.php?pid=751054}{Abhijit Banerjee} and \href{https://academictree.org/econ/tree.php?pid=175243}{Michael Kremer}. Her third adviser was \href{https://academictree.org/econ/tree.php?pid=743440}{Joshua Angrist}, the 2021 winner, who's adviser was \href{https://academictree.org/econ/tree.php?pid=167344}{David Card}, who also won in 2021. Card is a great-great-grandstudent of Modigliani. Banerjee is a student of Maskin. This makes Duflo a fourth-generation Nobelist. Kremer is a great-grandstudent of Schultz, via Zvi Griliches and Robert Barro.

\href{https://academictree.org/econ/tree.php?pid=181672}{Paul Milgrom} shared the Nobel prize with his advisor, \href{https://academictree.org/math/tree.php?pid=646344}{Robert B. Wilson}. Milgrom is the third of Wilson's students to win the Nobel prize, after Holmstr{\"o}m and Roth.

The 2021 Nobelists Angrist and Card were added to the tree in 2019. They are the fourth professor-student pair to win the Nobel prize. The third 2021 Nobelist, \href{https://academictree.org/econ/tree.php?pid=250252}{Guido Imbens}, is a grandstudent of Frank Hahn and so descends from Ely, and a great-great-grandstudent of Keynes.

There are 376 men in the graph, and 5 women. There are 85 male Nobelists, 2 female.

\subsection{The Nobel network over time}
\begin{landscape}
\begin{figure}[p]
    \centering
    \includegraphics[width=1.2\textwidth]{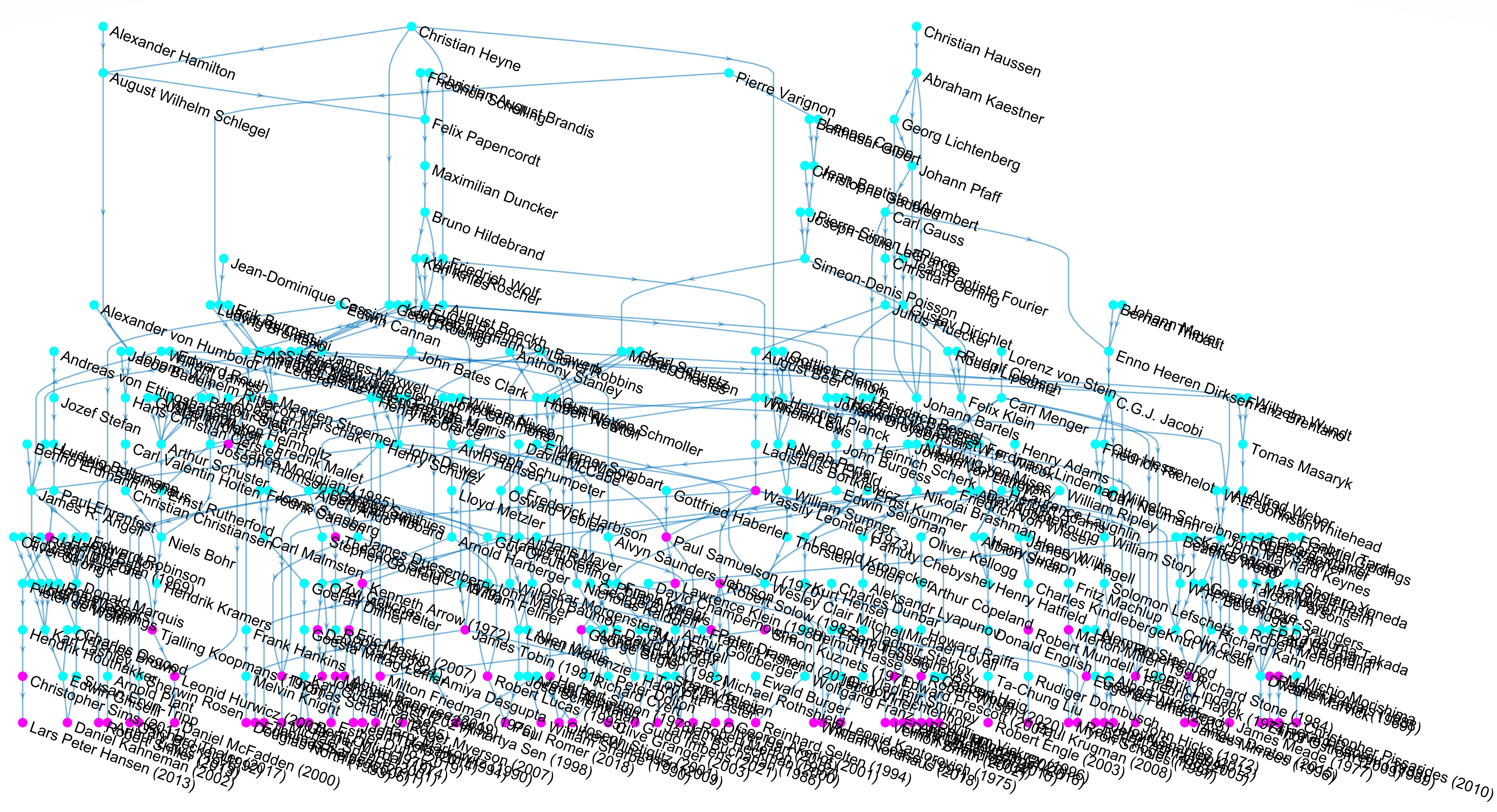}
    \caption{The complete professor-student network of Nobel Prize laureates in economics.}
    \begin{figurenotes}
        Nobelists are marked in magenta.
    \end{figurenotes}
    \label{fig:complete}
\end{figure}
\end{landscape}

Figure \ref{fig:complete} shows the complete network for 2021; there is also a \href{https://www.youtube.com/watch?v=UNEIZ2hbaqk}{video} of all networks since 1969. The tree starts with Alexander Hamilton (of musical fame), Christian Heyne (a philosopher) and Christian Hausen (a mathematician). Modigliani and Leontief are the Nobelists with the largest number of descendant generations. The Nobelists who have yet to produce Nobel offspring are at the bottom of the graph.

\begin{figure}[p]
    \includegraphics[width=\textwidth]{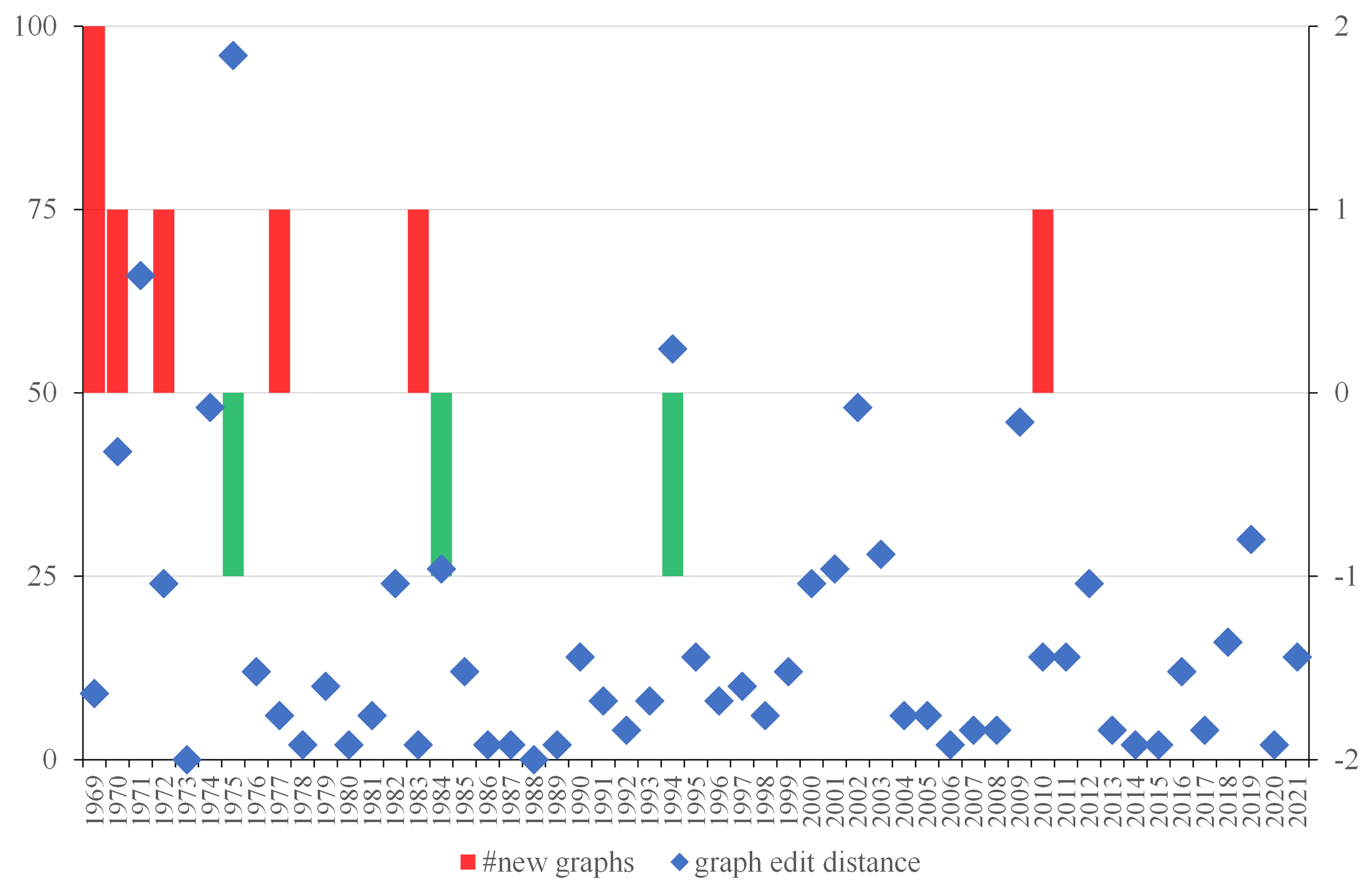}
    \caption{Graph edit distance and change in the number of disjoint graphs.}
    \label{fig:dgraph}
\end{figure}

Figure \ref{fig:dgraph} quantifies the changes over time. We start with two disjoint graphs. At five occasions, a new, disjoint graph is added. At three occasions, previously disjoint graphs are joined. At the end, there are four disjoint graphs: Frisch-Haavelmo; Allais-Debreu; Pissarides; and all other Nobelists. The final network is messy, as professors group and regroup to advise different students, and professors team up with their students and grandstudents to teach a new generation\textemdash but the key point is that Nobelists are connected.

Figure \ref{fig:dgraph} shows that the graph edit distance\textemdash the number of new nodes and edges\textemdash is often small. In 28 out of 53 years, the distance is 10 or less. This is a different way of revealing the same information: New Nobelists are closely related to previous Nobelists. The largest change in the network was in 1975, when 96 edits were needed to include Kantorovich and Koopmans and connect them to the main graph. 66 edits where needed to connect Kuznets to Samuelson in 1971. 56 edits were needed in 1994 for (Harsanyi,) Nash and Selten. Kahneman (and Smith) in 2002 needed 48 edits, and Ostrom (and Williamson) 40 edits in 2009. These results may strike us as odd, but the current reader sees the earlier Nobel laureates as the ultimate insiders. They were not at the time.

Furthermore, as shown above, Kahneman and Ostrom are not as alien to the economics profession as sometimes claimed. Both studied core issues in economics\textemdash rationality and public goods, respectively\textemdash but from a different perspective\textemdash psychology and political science, respectively. But they share their academic lineage with many of the great economists.

TOT HIER

Table \ref{tab:race} shows the central-most professors in the network, over time.\footnote{See \citet{Waumans2016} for the evolution over time of the citation centrality of academic papers.} On both distance measures\textemdash harmonic, see Equation (\ref{eq:harmmean}), and arithmetic, see Equation (\ref{eq:arithmean})\textemdash Karl Knies is the most central person at the moment. Using the arithmetic mean distance, which emphasizes connectivity over proximity, he has been since 1981. Using the harmonic mean distance, which emphasizes proximity over connectivity, Knies and Leontief have frequently swapped rank, with Robert Shiller pushing Knies into the lead in 2013, a position weakened by Oliver Hart in 2016 but strengthened by Richard Thaler in 2017. Others, including Henry Schultz and Josef Schumpeter, were particularly central in earlier years.

\begin{table}[p]
\caption{Central-most professors over time.}
\begin{tabular}{lll}
& arithmetic & harmonic \\ 
Knies & 1: 1981-2021 & 1: 2019-2021, 2013-2017, 2007-2009, 1982-2004 \\ 
& 2: - & 2: 2018, 2016, 2010-2012, 2005-2005, 1981\\
& 3: - & 3: 1980\\
Leontief & 1: 1970-1972 & 1: 2018, 2010-2012, 2005-2006, 1970-1972\\
& 2: - & 2: 2019-2021, 2017, 2015, 2007-2009, 2004\\
& 3: 2004-2021, 1973 & 3: 2002-2003, 1973\\
Schultz & 1: 1978-1980 & 1: 1978-1979\\
& 3: - & 3: 1995, 1980\\
Bawerk & 3: 1990-1992, 1985-1986, 1981 & 3: 1990-1992, 1985-1986\\
Sombart & 1: 1973 & 1: -\\
& 2: - & 2: 1981\\
& 3: 1987-1989, 1974-1976 & 3: 1987-1989\\
Schumpeter & 1: 1970-1972 & 1: 1981, 1970-1972\\
& 2: 1981-1983 & 2: 1982-1983\\
& 3: 1984, 1973 & 3: 1984-1986, 1973\\
Mises & 1: 1980, 1974-1976, 1972 & 1: 1972-1980\\
& 2: 1977 & 2: 1981\\
& 3: 1978-1979, 1973 & 3: -\\
Menger & 2: 1974-1976 & 2: 1980, 1974-1976\\
& 3: 1977 & 3: 1977\\
Cassel & 1: 1977-1980 & 1: 1977-1979\\
& 3: - & 3: 1980
\label{tab:race}
\end{tabular}
\begin{tablenotes}
Shown are the ranks of the most central professors of economics, as measured by their (arithmetic or harmonic) average distance to Nobelists. Leaders in the period 1969-1978 are omitted. Knies' professor, Bruno Hildebrand, and grandprofessor, Maximilian Duncker, are also left out.
\end{tablenotes}
\end{table}

Table \ref{tab:rank} shows the current top 10. The key people are Knies and Leontief. Their professors and grandprofessors are included because they too are academic ancestors of Nobel Laureates, even though they did not contribute (much) to the discipline. Schumpeter is included as the professor of Samuelson and Tobin and the grandprofessor of Schelling. Heyne connects three branches of the main graph.

\begin{table}[p]
\caption{Central-most professors in 2021.}
\begin{tabular}{lcc}
& arithmetic & harmonic \\ 
1 & Knies & Knies\\
2 & Hildebrand & Leontief\\
3 & Leontief & Hildebrand\\
4 & Duncker & Sombart\\
5 & Sombart & Duncker\\
6 & Bawerk & Bawerk\\
7 & Papencordt & Bortkiewicz\\
8 & Heyne & Papencordt\\
9 & Bortkiewicz & Heyne\\
10 & Schlegel & Schultz
\label{tab:rank}
\end{tabular}
\begin{tablenotes}
Shown are the ten most central professors of economics, as measured by their (arithmetic or harmonic) average distance to Nobelists.
\end{tablenotes}
\end{table}

\begin{figure}[p]
    \centering
    \includegraphics[width=\textwidth]{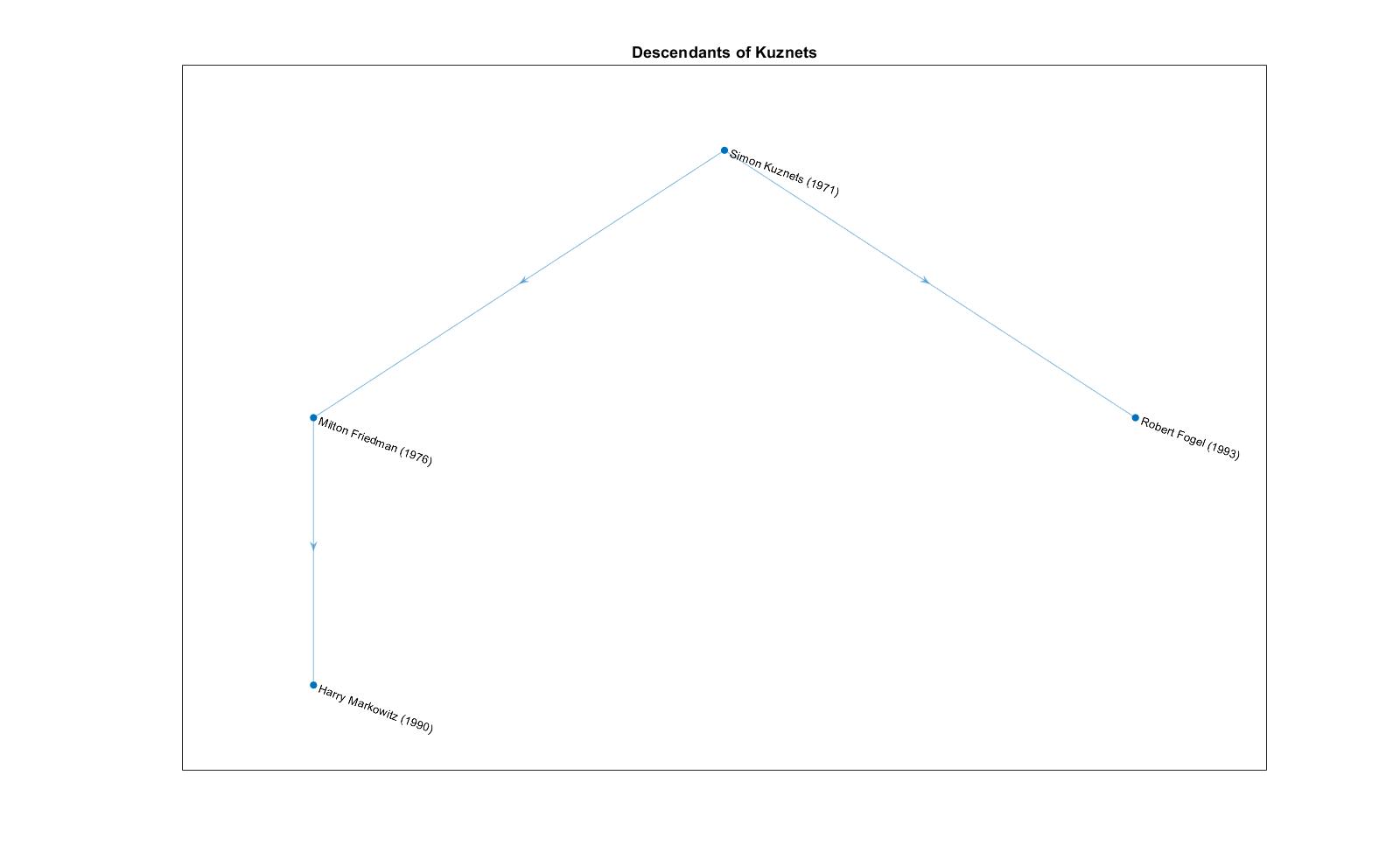}
    \includegraphics[width=\textwidth]{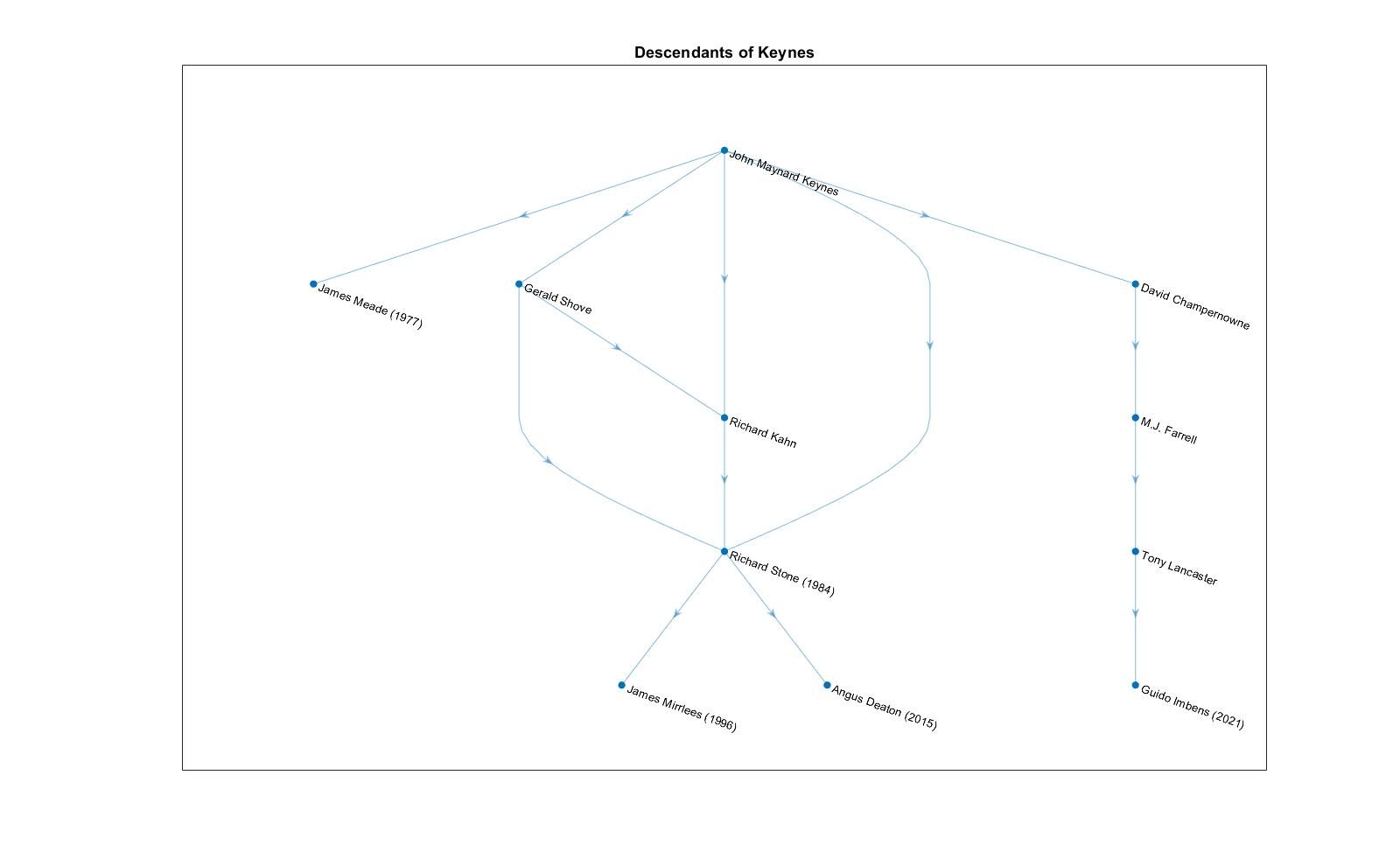}
    \caption{Selected subgraphs: Kuznets and Keynes.}
    \label{fig:tinbergen}
\end{figure}

\begin{figure}[p]
    \centering
    \includegraphics[width=\textwidth]{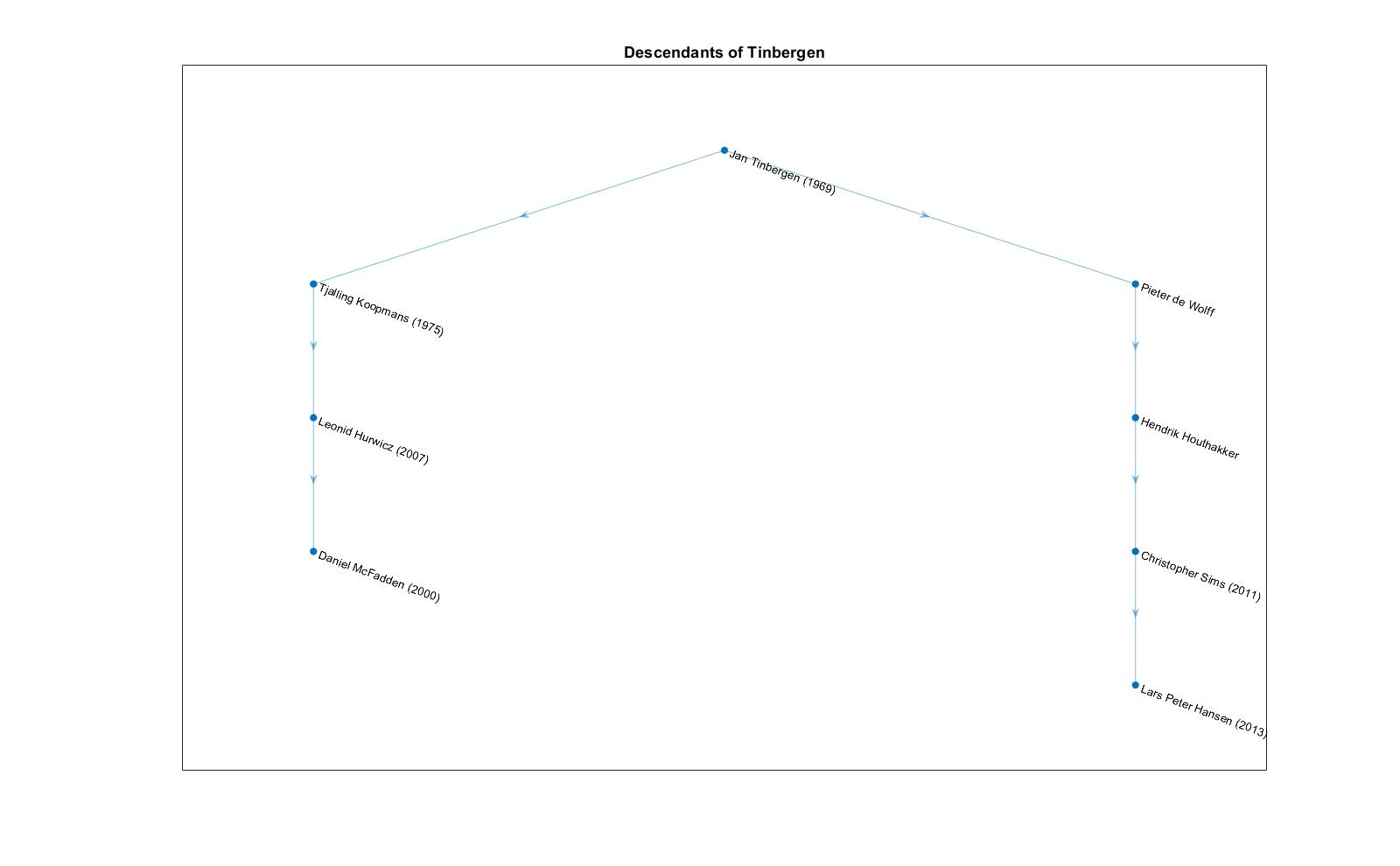}
    \includegraphics[width=\textwidth]{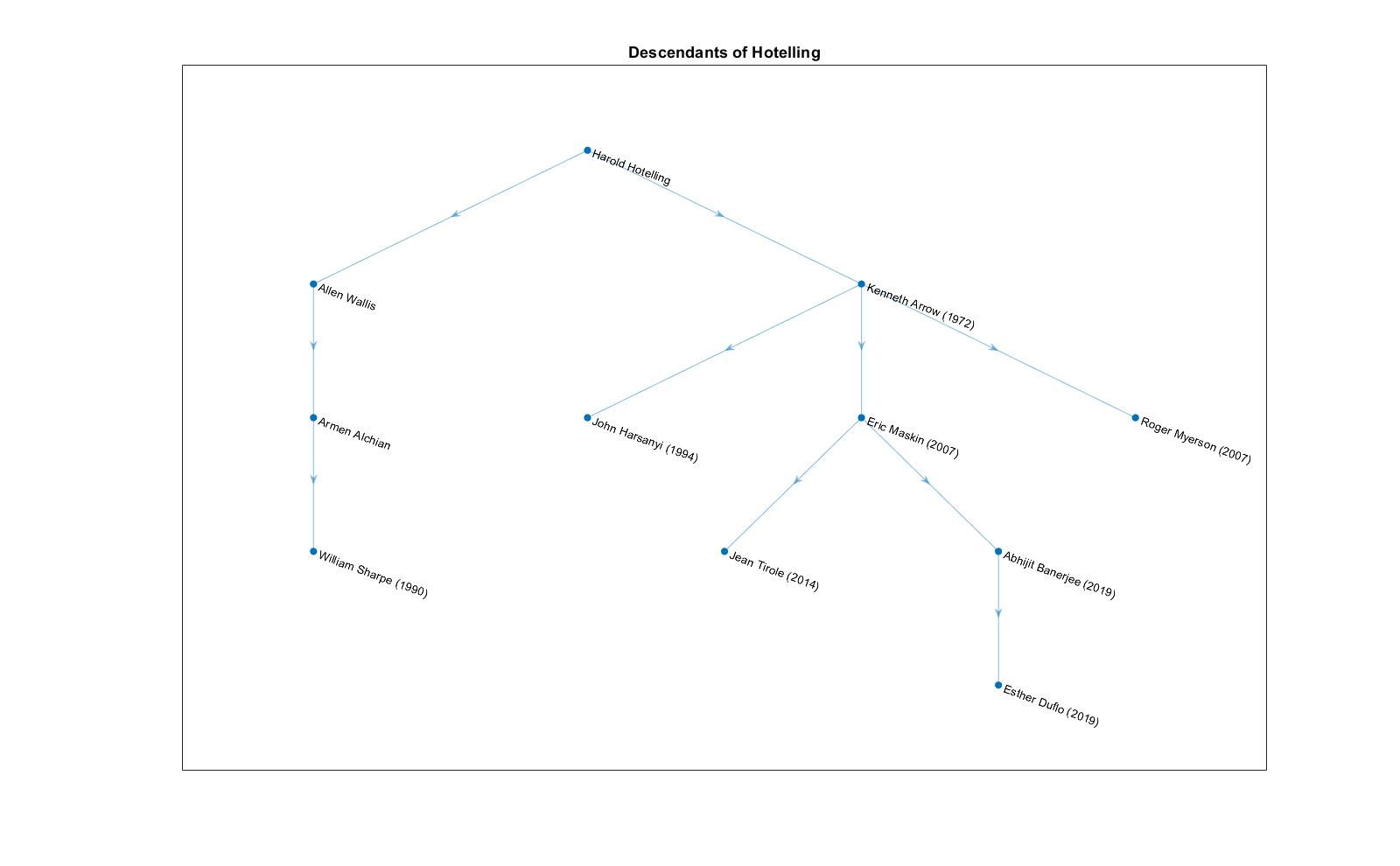}
    \caption{Selected subgraphs: Tinberger and Hotelling.}
    \label{fig:hotelling}
\end{figure}

\begin{figure}[p]
    \includegraphics[width=\textwidth]{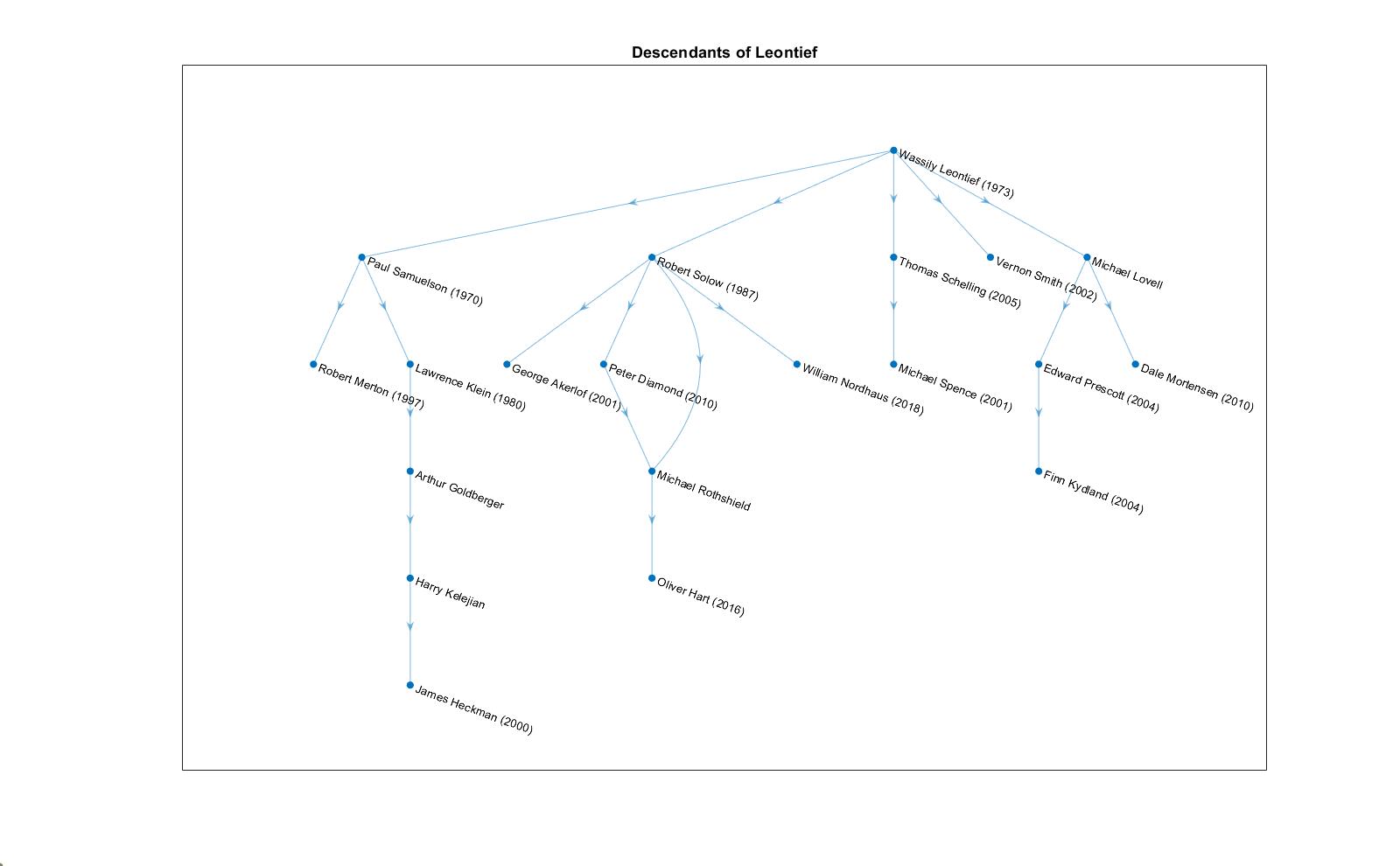}
    \includegraphics[width=\textwidth]{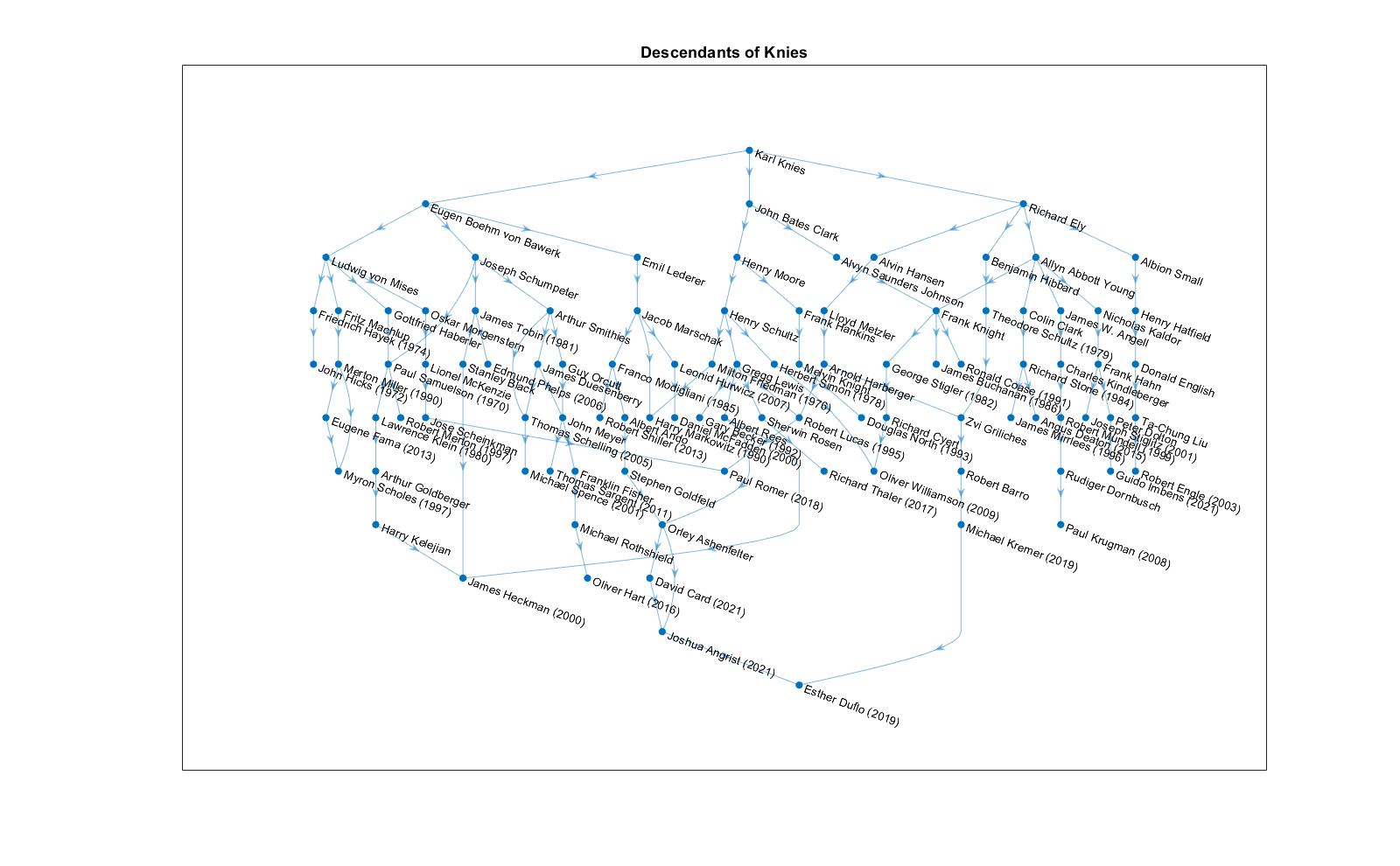}
    \caption{Selected subgraphs: Leontief and Knies.}
    \label{fig:knies}
\end{figure}

Figures \ref{fig:tinbergen} and \ref{fig:hotelling} shows selected subgraphs, for Kuznets (4 Nobelists), Keynes (5 Nobelists), Tinbergen (6 Nobelists) and Hotelling (8 Nobelists). These graphs are simple. The Hotelling graph shows the branching that make it hard to display made nodes. The Keynes graph shows that Stone was his student, grandstudent, and great-grandstudent. Figure \ref{fig:knies} shows the subgraphs for Knies (42 Nobelists, all distant) and Leontief (16 Nobelists, most close). The Leontief shows the key role he and his (students') students played in economics. The 1973 and 2000 Nobel prizes span six generations. The Knies graph again shows the complex relationships between Nobelists. The tree first splits between Bawerk, Clark and Ely and later rejoins.

\subsection{Sensitivity analysis}
Figure \ref{fig:complete} is based on the nearest common ancestor of a new Laureate and any of the previous ones. This algorithm does not necessarily lead to the same, minimum spanning graph if all Nobel Prizes were awarded in 2021. In order to test whether this affects centrality, I include all ancestors of all Laureates, appropriately cut-off at \href{https://academictree.org/philosophy/tree.php?pid=7045}{William of Ockham}.\footnote{The interested reader can trace Ockham's lineage to John the Baptist.} The expanded graph is quite a bit larger, with 1103 instead of 381 nodes. Centrality does not change much, however. The main difference is that \href{https://neurotree.org/neurotree/tree.php?pid=8563}{Jakob Thomasius} enters the top 10 of most central ancestors. Thomasius was a 17\textsuperscript{th} century philosopher at the University of Leipzig, and a common ancestor of common ancestors. The expanded graph contains many famous philosophers, theologists, jurists, physicists, biologists, and chemists\textemdash but no classical or pre-classical economists.

Some entries on \href{https://academictree.org/}{AcademicTree.org} are disputed and historical facts continue to be discovered. I focus on four individuals. Carl Menger supervised the \textit{Habilitation} of Eugen B\"{o}hm von Bawerk, and arranged for his studies with Knies and Hildebrand. This edge was added. Leonid Hurwicz fled Poland after obtaining his PhD from an unknown adviser in 1938. While von Hayek and von Mises provided shelter, they did not contribute to his education. Similarly, John Hicks joined Hayek and Robbins after finishing his education. These edges were removed. This would create a separate Hurwicz-McFadden tree. However, McFadden edited his AcademicTree entry to include Arrow, Uzawa (a student of Arrow) and Chipman (a student of Machlup). These edges were added. In his autobiography, Gary Becker stresses the importance of Director, Friedman, Savage and Schultz to his intellectual development. Alfred Marshall mentored John Maynard Keynes. These edges were added.\footnote{I am grateful to an anonymous referee for three of the six adjustments, two other referees for two others.}

These changes in the network lead to changes in the centrality of the people in the network. Koopmans and other people in the Tinbergen tree become less central. Friedman and Menger become more central. Karl Knies remains the central-most figure in the network.

\subsection{Universities}
\begin{figure}[h]
    \includegraphics[width=\textwidth]{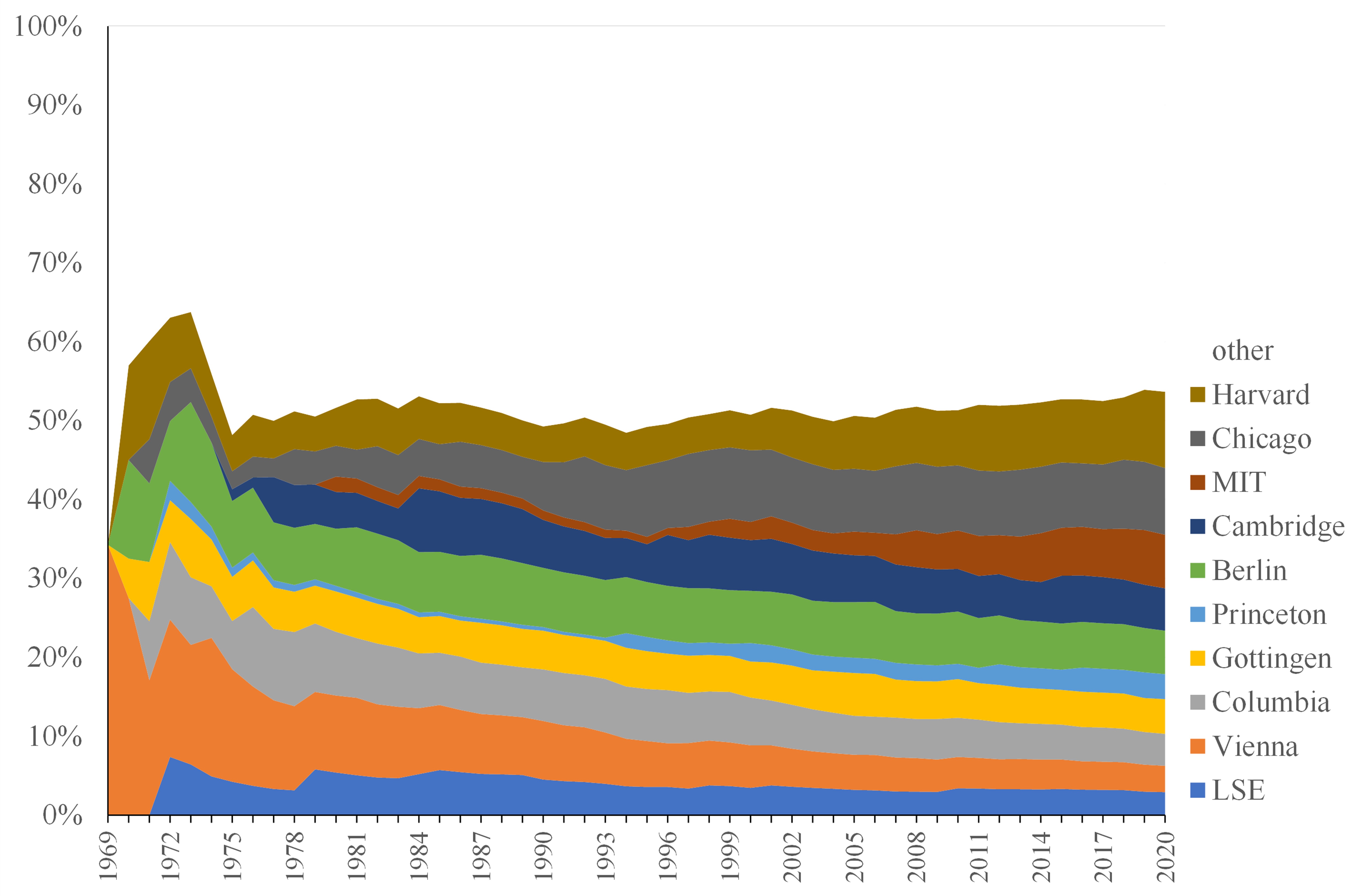}
    \caption{Universities that granted most degrees to Nobel Laureates and their professors.}
    \label{fig:uni}
\end{figure}

Figure \ref{fig:uni} shows the key places of learning, ranked by their centrality in 2020. A \href{https://www.youtube.com/watch?v=n1bfu8oyi6w}{video} shows location and centrality over time. I show total centrality: A university is granted one point if a Nobel Laureate obtained his PhD there. The university where his professor graduated is awarded half a point, with further ancestors' universities granted points in proportion to the arithmetic centrality measure (\ref{eq:arithmean}). Figure \ref{fig:uni} shows market share, as levels increase with time. In total, 61 universities are included but Figure \ref{fig:uni} only shows the top 10. Harvard comes out top, closely followed by Chicago and MIT. Berlin, G{\"o}ttingen and Vienna rank highly for their rich heritage rather than their current prowess.\footnote{Similarly, the historic roots of psychology can be found in Leipzig \citep{Rieber2001}.} Cambridge and the London School of Economics are the only UK institutes on the list, but highly ranked. Columbia and Princeton complete the top 10. Further down the list, Heidelberg bests Yale and Stanford. Paris ranks above Johns Hopkins and Leiden, which never had an economics department, ahead of Berkeley.

Figure \ref{fig:uni} thus reveals the dramatic shifts in the geography of teaching economics. When the generation of John Bates Clark studied economics, Heidelberg was the university to be and Knies the professor to consult.\footnote{\citet{Perlman2017} argues that German universities were ahead of the European competition because they were free from Church control and, after the Bismarck reforms, local state control. \citet{Balabkins2017} adds that American students of economics preferred Germany to Great Britain because the latter emphasized theory and a doctrinal adherence to laissez-faire.} Young economists today want to study with Knies' great-great-great-grandstudents, who occupy the halls of the top universities in the USA.

\section{Nobel candidates}
Clarivate's list of citation laureates includes many of world's leading economists, although there are some surprising omissions too. All would be worthy Nobelists, and I suspect many will be. Indeed, Angrist, Card, Deaton, Engle, Fama, Granger, Hansen, Hart, Holmstr{\"o}m, Kahneman, Krugman, Milgrom, Nordhaus, Romer, Sargent, Shiller, Sims, Thaler, Tirole, Williamson, and Wilson were citation laureates before winning the Nobel Prize. In Appendix \ref{app:clarivate}, I discuss the other citation laureates grouped by their place in the Nobel family tree, starting with the closest relatives.

\subsection{The Nobel network in the future}
The future network of Nobelists is likely to look much like the current network, as many candidates are closely related to previous winners and others most distantly. There are only three women among the candidates. New graphs may emerge around Gorman and Sargan, with the Gormanites more readily connected to the main tree than the Sarganites.

\begin{figure}[p]
    \includegraphics[width=\textwidth]{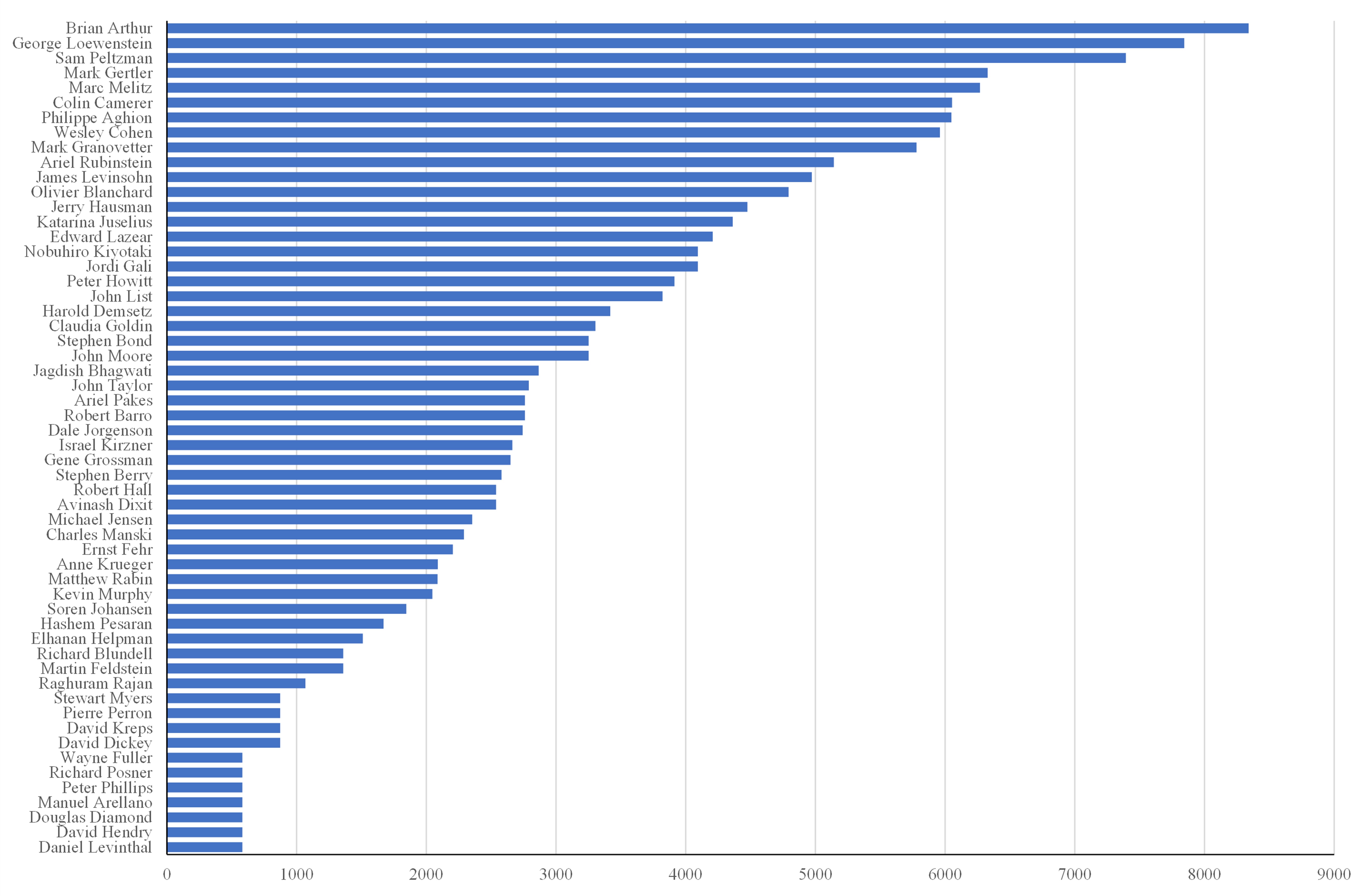}
    \includegraphics[width=\textwidth]{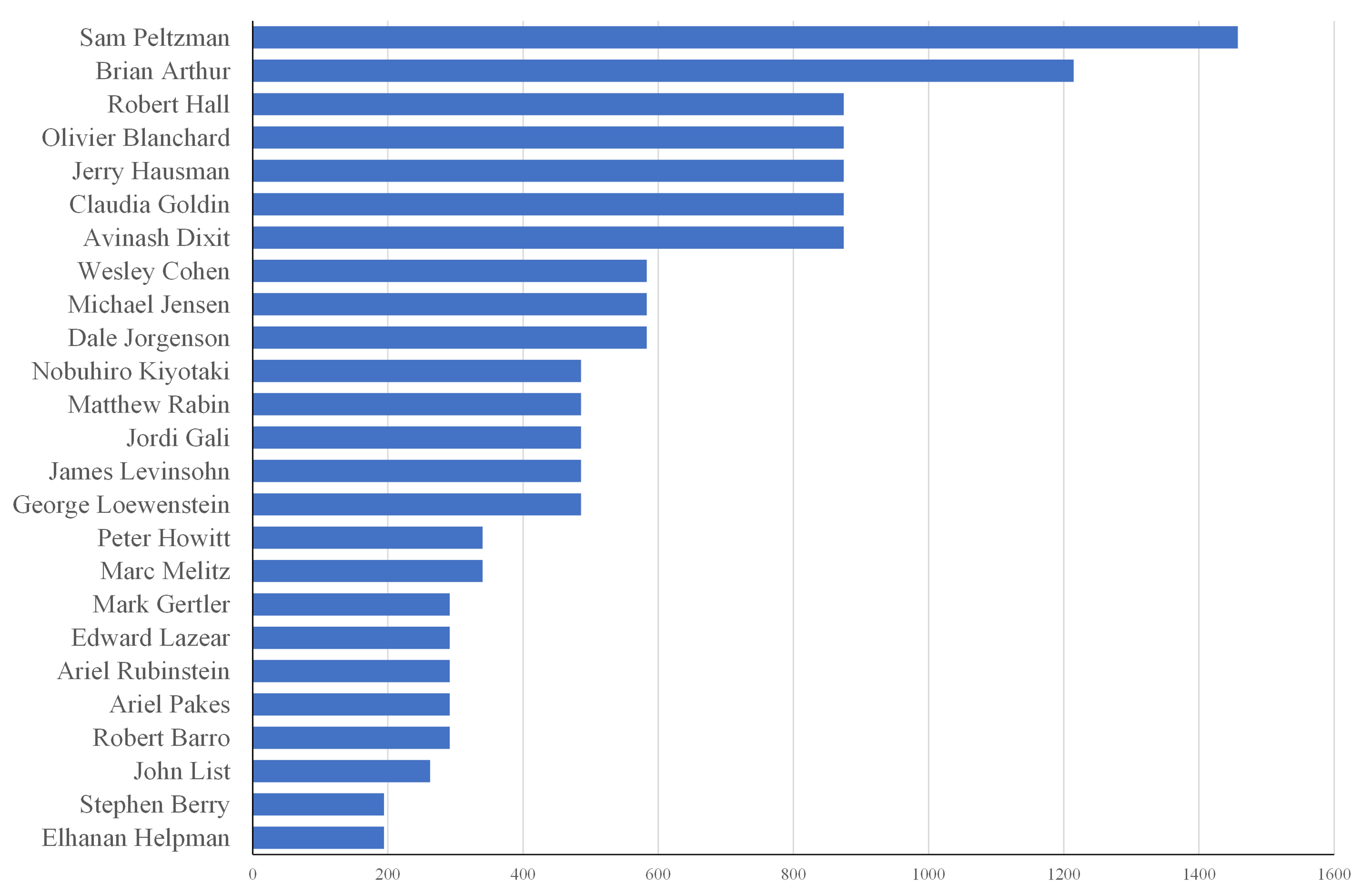}
    \caption{Harmonic mean distance of Nobel candidates to the 2021 network (top panel) and to the Nobelists (bottom panel).}
    \label{fig:cand2network}
\end{figure}

The top graph in Figure \ref{fig:cand2network} shows the harmonic average distance from the Nobel candidates to the network in 2021. This is a measure of how central the candidate would be, should (s)he win. Brian Arthur, George Loewenstein, and Sam Peltzman are closest to the current network, Douglas Diamond, David Hendry, and Daniel Levinthal the furthest removed.

The bottom graph in Figure \ref{fig:cand2network} shows the harmonic average distance from the Nobel candidates to the current Nobel laureates. Sam Peltzman now comes top, followed by Brian Arthur, and Blanchard, Dixit, Goldin, Hausman, and Hall in shared third place. 31 of the 56 candidates do not have a Nobel ancestor.

\begin{figure}[h]
    \includegraphics[width=\textwidth]{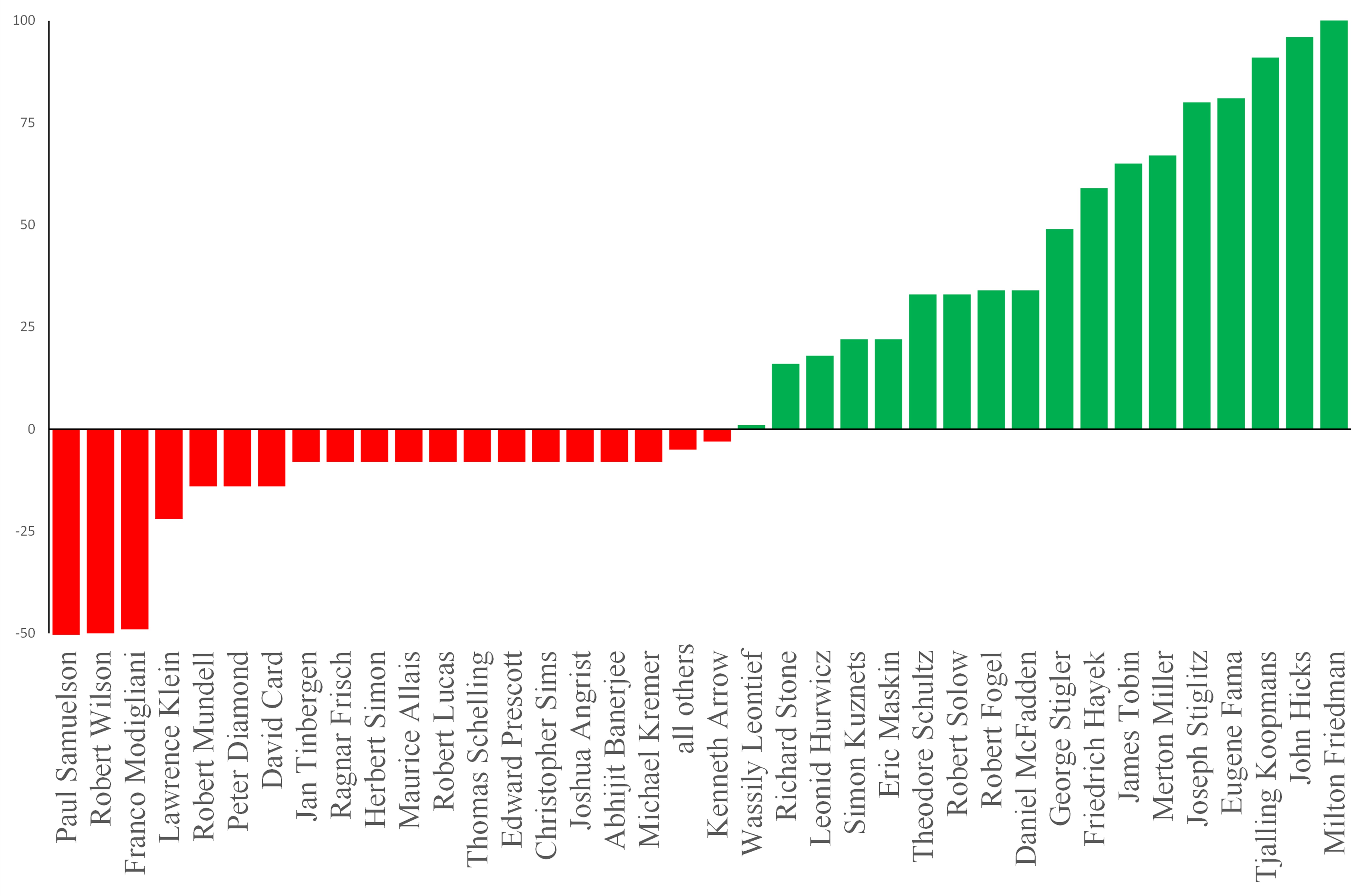}
    \caption{Change in the harmonic centrality rank of selected Nobelists if all candidates win.}
    \label{fig:nobel2cand}
\end{figure}

Figure \ref{fig:nobel2cand} shows the change in the central ranking, based on the harmonic average distance, of Nobel Laureates should all candidates win. Milton Friedman, John Hicks and Tjalling Koopmans are closest to the candidates. Most Nobelists are not directly connected, but drop five places as others are propelled to greater centrality. Paul Samuelson, Robert Wilson and Franco Modigliani are furthest removed from the candidates, primarily because most of their Nobel-worthy students have won.

\section{Discussion and conclusion}
I show that the vast majority of winners of the Nobel Memorial Prize in Economic Sciences are connected in a network of professor-student relationships, and that this is likely to remain the case for years to come. The underlying data can be inspected and extended by anyone. A closer look at the wider network quickly dispels the notion of aggrieved economists, or their champions, that they did not want for lack of connections.

The two central figures in the Nobel network are Karl Knies and Wassily Leontief. Leontief needs no introduction \citep{Samuelson2004, Baumol2009, Dietzenbacher2004, Debresson2004, Duchin1995, Missemer2020}. Who was Karl Knies? Having to raise this question indicates that his contributions to economics lie primarily in taking several young men\textemdash Herbert Baxter Adams, Eugen B{\"o}hm von Bawerk, John Bates Clark, Richard T. Ely, Richmond Mayo-Smith and Edwin Seligman\textemdash under his wings. Little has been written about Knies \citep{Kobayashi2002, Schefold2008, Fullerton1998, Trautwein2005, Yagi2005}, and there is no theorem, curve, lecture or medal named after him. Knies was a professor of public policy at the University of Freiburg from 1855 to 1865, and a professor of economics at the University of Heidelberg for the thirty years after, succeeded by Max Weber. He published on statistics, utility, political economy, railroads, and money and credit. Knies was part of the German Historical School, who argued for empiricism over theory, and thus at the losing end of the \textit{Methodenstreit} \citep{Louzek2011, Maki1997}. \citet{Ely1938} does not divulge much detail, but notes that Knies thought he did not receive the credit he deserved.\footnote{The economics department at the University of Heidelberg is named after Alfred Weber, Max' brother and successor.} Knies was thus somewhat of an outlier, arguing for empirical analysis as the discipline of economics was turning towards theory. Knies would have welcomed the credibility revolution and the 2021 Nobel prize, shared by two of his descendants.

Leontief is best known as the father of input-output analysis \citep{Polenske2004}, a tool we now rarely see used at the cutting edge of economics. The main prize for heterodox economists is named after Leontief. The central role of Knies and Leontief in economic orthodoxy may thus come as a surprise\textemdash unless you realize that economics is at its best when the received wisdom is challenged \citep{Roth2018}.

If the central professors surprise, so do some of the key degree-granting universities: Berlin, G{\"o}ttingen, Vienna and Heidelberg are not typically listed among the leading centres of research and education in economics. However, these universities did educate many of the Nobelists' professors and their professors. Another surprising finding, the great economists of the past are not included in the family tree, but great scholars in just about any other discipline are (see discussion in the introduction). Putting these two findings together, it appears that during the first half of the 20\textsuperscript{th} century, economists in Boston, Chicago and New York combined the ideas of earlier economists with the teaching methods of other disciplines and countries and so created the economics powerhouses we see today. Refugees from the Nazis\textemdash Haberler, Hayek, Hurwicz, Machlup, Marschak, von Mises, Morgenstern\textemdash played a key role in this, and ones who fled the Bolsheviks\textemdash Domar, Kuznets, Leontief and Marschak. See \citet{Scherer2000} and \citet{Hagemann2011} for an account.\footnote{See \citet{Waldinger2010} and \citet{Waldinger2012}, and references therein, for the impact of the Nazis on physics, chemistry and mathematics.} \citet[p.43]{Ely1938} writes that "[y]ou learn here [in Germany], and only here, how to do independent, real scientific work." \citet[p.40]{Ely1938} notes that "[a]lthough [\href{https://neurotree.org/neurotree/tree.php?pid=751775}{Johannes Conrad} of the University of Halle] was not a great original thinker he was a splendid teacher", emphasizing the then disconnect between those who taught economics and those who moved its frontier.

The Nobel network has a gender-ratio that is even more biased towards men than the \href{https://ideas.repec.org/top/female.html}{economics discipline as a whole}.

The strong concentration of Nobelists around a small number of professors and universities is reminiscent of the strong concentration of top economists, top departments and top journals \citep{Hamermesh2013, Torgler2013}.

I focus on professor-student relationships. There are, of course, many other types of interactions, including citation, co-supervision, shared supervisors, co-authoring\textemdash \citet{ArrowDebreu1954}, for example\textemdash informal advice\textemdash John Maynard Keynes was the most senior figure in the Cambridge Circus, to which Joan Robinson also belonged\textemdash and seminars\textemdash in his autobiography, Herb \citet{Simon1996} fondly reminisces the Cowles seminars, attended by in-house staff Marschak, Koopmans, Lange, Arrow, Klein, Hurwicz and Debreu and regular visitors Modigliani, Stigler, Friedman, Frisch and Haavelmo. Simon also describes a mind-shifting lunch with Carl Menger. \citet[p. 77-78]{Torgler2013} document, in the words of Becker, Coase, Friedman and Samuelson, the contribution of Frank Knight to their intellectual development, in the absence of a formal mentoring relationship. Adding such relationships would only make the network more dense. It may well lead to a shift in network centrality. Data on informal contacts are, of course, much harder to obtain. \citet{Emmett2009} demonstrates the power of such an analysis, and the effort required. Such trees are probably best limited to a certain department, as Emmett does, or a subfield of economics.

The descriptive analysis in this paper would be served with additional data on selected ancestors, such as discipline, ethnicity, religion. I identify a set of people who trained exceptional students, a set of people who did exceptional research, and the intersection of these sets. This should form a stepping stone to an analysis of what makes someone an exceptional researcher, an exceptional professor, or both. Contrasting the citation tree with the supervision tree would distinguish the two.

The measure, here defined, on incloseness to either the Nobel network or Nobel laureates can be calculated for individual researchers, departments, or editorial boards\textemdash and should be tested as a predictor for academic performance, career advancement, and the probability of winning the John Bates Clark Medal or indeed the Nobel Memorial Prize in Economic Sciences. Family trees could be constructed for similarly prestigious prizes in other disciplines to test whether tight clustering is unique to economics. Links between family trees\textemdash via Tinbergen to the Nobel Prize in Physics, via Nash to the Abel Prize, and via Knies to history, sociology and political science\textemdash would be worthy of study too. All that is deferred to future research.

\bibliographystyle{aea}
\bibliography{nobel}

\appendix

\section{Clarivate's citation laureates}
\label{app:clarivate}
\subsection{Students and academic siblings}
\href{https://academictree.org/econ/tree.php?pid=830522}{Brian Arthur} was a student of McFadden. He would be the first fifth-generation Nobelist. 

\href{https://academictree.org/econ/tree.php?pid=742311}{Olivier Blanchard}, \href{https://academictree.org/econ/tree.php?pid=188231}{Avinash Dixit} and  \href{https://academictree.org/econ/tree.php?pid=181660}{Robert Hall} were all students of Solow. So are Anthony Atkinson, Martin Weitzman and Halbert White, who died before winning.

\href{https://academictree.org/econ/tree.php?pid=175233}{Claudia Goldin} was a student of Fogel. \href{https://academictree.org/econ/tree.php?pid=722576}{Jerry Hausman} was Mirrlees' student. \href{https://academictree.org/econ/tree.php?pid=175240}{Dale Jorgenson} was Leontief's student, \href{https://academictree.org/econ/tree.php?pid=743485}{Michael Jensen} a student of Miller, \href{https://academictree.org/econ/tree.php?pid=516547}{Sam Peltzman} a student of Friedman and Stigler.

\href{https://academictree.org/econ/tree.php?pid=181088}{Harold Demsetz}, a student of \href{https://neurotree.org/neurotree/tree.php?pid=738966}{Frank Knight}, is a sibling of Buchanan, Coase and Stigler. The late Gordon Tullock was a grandstudent of Knight.

\href{https://academictree.org/econ/tree.php?pid=743440}{Joshua Angrist} and \href{https://academictree.org/econ/tree.php?pid=167344&fontsize=1&pnodecount=4&cnodecount=2}{David Card} were students of \href{https://academictree.org/econ/tree.php?pid=188179}{Orley Ashenfelter} and so descend from Lewis and Modigliani. Angrist was advised by Card and advised Duflo. An Angrist-Card Nobel prize would make Duflo a third-generation Nobelist; she is already a fourth-generation one.

\subsection{Grandstudents and academic cousins}
\href{https://academictree.org/econ/tree.php?pid=743450}{Jordi Gali} and \href{https://academictree.org/econ/tree.php?pid=188202}{Nobuhiro Kiyotaki} are grandstudents of Solow, and Marc Melitz a great-grandstudent. All are candidate fourth-generation Nobelists, conditional on a Blanchard Nobel Prize.

\href{https://academictree.org/econ/tree.php?pid=542983}{Wesley Cohen} is a grandstudent of Schelling, Stiglitz and Tobin. \href{https://academictree.org/econ/tree.php?pid=543016}{George Loewenstein} is a grandstudent of Stiglitz, \href{https://academictree.org/econ/tree.php?pid=409581}{Mark Gertler} Tobin's.

\href{https://academictree.org/econ/tree.php?pid=743447}{Kenneth French} is Fama's grandstudent.

\href{https://academictree.org/econ/tree.php?pid=175255}{Matthew Rabin} is a grandstudent of Maskin, \href{https://academictree.org/econ/tree.php?pid=409537}{Ariel Rubinstein} of Arrow. The late Armen Alchian was Arrow's academic nephew.

\href{https://academictree.org/econ/tree.php?pid=415934}{Philippe Aghion} is a grandstudent of \href{https://academictree.org/econ/tree.php?pid=738543}{Lionel McKenzie}, and so descends from Baumol, Cannan and Knies. Aghion further descends from Poisson. He is an academic cousin of Romer.

\subsection{Distant relations}
\href{https://academictree.org/econ/tree.php?pid=175217}{Robert Barro}, \href{https://academictree.org/econ/tree.php?pid=552805}{Edward Lazear} and \href{https://academictree.org/econ/tree.php?pid=175253}{Ariel Pakes} were students of \href{https://academictree.org/econ/tree.php?pid=717822}{Zvi Griliches}, and \href{https://academictree.org/econ/tree.php?pid=830516}{Steven Berry} a grandstudent. They so descend from Ely. \href{https://academictree.org/econ/tree.php?pid=183273}{Jagdish Bhagwati}, \href{https://academictree.org/econ/tree.php?pid=188194}{Gene Grossman}, \href{https://academictree.org/econ/tree.php?pid=519783}{James Levinsohn} and \href{https://academictree.org/econ/tree.php?pid=175249}{Marc Melitz} also descend from Ely, via \href{https://academictree.org/econ/tree.php?pid=743441}{Charles Kindleberger}. \href{https://academictree.org/econ/tree.php?pid=181694}{Anne Krueger} is one of three woman among the Nobel candidates. Ely is her great-great-grandprofessor.

\href{https://neurotree.org/neurotree/tree.php?pid=5264}{Colin Camerer} is a great-great-great-grandstudent of \href{https://academictree.org/psych/tree.php?pid=124}{John Dewey}, Kuznets' grandprofessor. \href{https://neurotree.org/neurotree/tree.php?pid=22376}{Ernst Fehr} descends from \href{https://neurotree.org/neurotree/tree.php?pid=743726}{Lujo Brentano}, Marschak's grandprofessor.

\href{https://academictree.org/sociology/tree.php?pid=375462}{Mark Granovetter} is a great-grandstudent of \href{https://academictree.org/physics/tree.php?pid=1943&pnodecount=2&cnodecount=3&fontsize=1}{Niels Bohr}, who was Koopmans' grandprofessor. \href{https://academictree.org/econ/tree.php?pid=830517}{Katarina Juselius} descends from another one of Koopmans' grandprofessors, \href{https://academictree.org/physics/tree.php?pid=23346}{Paul Ehrenfest}. \href{https://academictree.org/econ/tree.php?pid=186314}{Peter Howitt} is great-grandstudent of Koopmans. \href{https://academictree.org/econ/tree.php?pid=175238}{Elhanan Helpman} descends from Ehrenfest via \href{https://academictree.org/econ/tree.php?pid=737349}{Houthakker}.

\href{https://academictree.org/econ/tree.php?pid=743486}{Israel Kirzner} was von Mises' student. \href{https://neurotree.org/neurotree/tree.php?pid=743534}{John List} also descends from von Mises, and from Hayek, Hicks and Keynes. \href{https://academictree.org/econ/tree.php?pid=743505}{Hashem Pesaran} descends from \href{https://academictree.org/econ/tree.php?pid=22201}{Keynes}. 

\href{https://academictree.org/math/tree.php?pid=379583}{Charles Manski} was a student of \href{https://academictree.org/econ/tree.php?pid=738891}{Franklin Fisher}, sharing an ancestry with other prominent economists, including several Nobelists. \href{https://academictree.org/econ/tree.php?pid=152661}{Kevin Myers} is a great-grandstudent of Schultz.

\href{https://academictree.org/econ/tree.php?pid=743397}{John Taylor} descends from \href{https://academictree.org/psych/tree.php?pid=94687&pnodecount=4&cnodecount=4&fontsize=1}{Everett Lindquist}, and is thus related to Amos Tversky and via  \href{https://academictree.org/math/tree.php?pid=207484}{Henry Rietz} to Gauss, Poisson and many Nobelists. 

\subsection{New trees}
\href{https://academictree.org/econ/tree.php?pid=731843}{Manuel Arellano}, \href{https://neurotree.org/neurotree/tree.php?pid=743483}{David Hendry} and \href{https://academictree.org/econ/tree.php?pid=96820}{Peter Phillips} were students of \href{https://academictree.org/econ/tree.php?pid=737291}{Denis Sargan}, and \href{https://academictree.org/econ/tree.php?pid=96818}{Pierre Perron} a grandstudent. Sargan was a self-taught man, unconnected to the Nobel graph. There are other distinguished econometricians in his tree.

\href{https://academictree.org/econ/tree.php?pid=743778}{Richard Blundell}, \href{https://academictree.org/etree/tree.php?pid=830524}{Stephen Bond}. \href{https://academictree.org/econ/tree.php?pid=175226}{Martin Feldstein}, \href{https://academictree.org/econ/tree.php?pid=743891}{John Moore} and the prematurely deceased Alberto Alesina and Alan Krueger share a common ancestor in \href{https://academictree.org/econ/tree.php?pid=738897}{Terence Gorman}. There are other prominent economists in the same family. Gorman is not connected to other Nobelists. Bond and Moore are also grandstudents of Frank Hahn, academic cousins of Imbens.

\href{https://academictree.org/econ/tree.php?pid=743502}{Stewart Myers} and his student \href{https://academictree.org/econ/tree.php?pid=415647}{Raghuram Rajan} descend from \href{https://academictree.org/math/tree.php?pid=160726}{David Alhadeff}. This could constitute a disjoint graph: Alhadeff obtained his PhD from Harvard shortly after World War II.\footnote{The registrar argues that student-professor relationships are private information.} 

The late Stephen Ross\footnote{PhD, 1970, Harvard} and his student \href{https://academictree.org/econ/tree.php?pid=415569}{Douglas Diamond} may form a disjoint graph, as would \href{https://academictree.org/math/tree.php?pid=250244}{Wayne Fuller} and his student \href{https://academictree.org/math/tree.php?pid=226664}{David Dickey}. \href{https://academictree.org/econ/tree.php?pid=830521}{Soren Johansen}, \href{https://academictree.org/econ/tree.php?pid=815875}{David Kreps}, \href{https://academictree.org/computerscience/tree.php?pid=478795}{Daniel Levinthal} and \href{https://academictree.org/law/tree.php?pid=675752}{Richard Posner} would form their own trees.

\section{Matlab codes}
\begin{itemize}
    \item Matlab \href{https://github.com/rtol/AcademicAncestry}{code} to scrape the ancestry of your favourite economist from Academic Tree and build her family tree.
    \item Matlab \href{https://github.com/rtol/NobelNetwork}{code} to build and analyze the data used in this paper.
\end{itemize}

\end{document}